# Agile Research

Hamish Cunningham
http://gate.ac.uk/hamish/

Internet Memory Foundation,
45 ter rue de la Révolution,
F-93100 Montreuil, France
hamish.cunningham a-t
internetmemory d-o-t n-e-t

Department of Computer Science,
University of Sheffield,
Regent Court, 211 Portobello Street,
Sheffield S1 4DP, UK.

February 3, 2012

# Contents

# Chapter 1

# Introduction

This paper discusses the application of agile software development methods in software-based research environments.

# Chapter 2

# Agile Software Development

To start with the obvious, building software is hard. How do people cope with hard construction problems? They hire an architect, who draws a lot of complicated pictures (or, more likely these days, builds a complex computer model), and, for big projects, builds a small scale prototype to give the commissioning customer the feel of the thing. This design process iterates until all appears to be well, and off we go to the cement pourers and the concrete reinforcers.

Software engineering, both as craft and later as discipline, began by making a similar analogy, and so adopted a commission-design-build conception of the ideal software development process. By the 1980s mature methods used techniques like flow charts, entity-relationship modelling and structure diagrams to capture the architectural team's vision of the system to be built (e.g. [Yourdon 89]). Later incarnations merged the data structures with the algorithms and created *object-oriented* design (e.g. [Booch 94]). These modelling approaches were then used to produce as comprehensive a design as possible, which was then turned over to implementation teams lower down the food chain (the skilled work was concentrated in architectural hands while the programmer's work taken to be almost as routine as bricklaying). Because of the step-by-step procession from high to low these methods were often called *waterfall* development.

And everyone lived happily ever after. Well, not really. The problem with the architectural metaphor for software engineering is that software and the contexts within which it is constructed have turned out to be rather unlike houses and bridges, and more like evolving viruses or cantankerous animals. This is due of a number of factors, including:

- The act of building software changes the needs and the ideas of the people who are going to use it. (This is true of buildings too – see [Brand 94] – but fewer people expect to be able to move the walls around every week or two.)

- The working systems that software is meant to support or replace are often so complex that no single person possesses a comprehensive understanding of them.

- Software increasingly sits at the bleeding edge of technology's interface with society, and the markets that condition which parts of this interface expand and contract are chaotic and fast-moving. In other words the context is both a strong determinant of success and extremely dynamic.

- Verification of the correct working of software is an extremely difficult problem to solve theoretically. Whereas we can be quite confident of the properties of a wall or joist built to certain well-established principles, the equivalent level of certainty with respect to software is much more difficult to achieve (and in the general case can only be approached via testing and not theoretical proofs[1]).

Therefore it has become a truism that the best way to build a particular piece of software is to have built a very similar one before [Frederick P. Brooks 78] (hence the famous phrase *'build one to throw away'*). Obviously this is often impossible, and especially so in an R&D context. The waterfall methods have failed[2] most often in highly dynamic contexts; the piles of design artefacts can end up out-of-date before they are finished, and the synchronisation of design and code a never-ending process which makes the nature of the executable system ever more opaque.

Over the last decade or so a new family of software engineering methods have arisen based on the types of insight sketched above: the *agile methods*. Below we discuss what we may usefully learn from this experience for our research work, beginning with a short summary of two of the most popular members of the agile family, Scrum and XP (eXtreme Programming).

## 2.1 Agile Teams and Scrum

In their book *Agile Software Development with Scrum* Ken Schwaber and Mike Beedle [Schwaber & Beedle 01] recall visiting a DuPont factory and talking with process engineers engaged in managing complex chemical processes. They described the types of activity they routinely encountered in software development and the types of (non-agile) design methods in use. The DuPont engineers laughed: they couldn't see how the methods matched up with the process because the methods were, from their point-of-view, appropriate only in much more defined and predictable circumstances. They suggested that for the chaotic and

---

[1] Automatic verification is a lively area of research but is still quite a distance from universal applicability.

[2] How to construct a failing software project: option 1: work for the UK government; option 2: define your plan as:

- first we (highly-paid developers) decide what to build (requirements)
- then we (middle-level developers) decide how to build it (design)
- then we (lower-level developers) implement it (coding)
- then we (support staff) deploy it (disaster)

A research equivalent: I hereby promise to invent The Next Big Thing at 3.27 PM on the 23rd of February 2010 in order to produce Deliverable D2,345 revision 63.

fast-moving software world a different approach was more appropriate, based on constant readjustment driven by empirical metrics. Scrum evolved as a response to these insights, and became

> ...a process skeleton that includes a set of practices and predefined roles. The main roles in Scrum are the ScrumMaster who maintains the processes and works similar to a project manager, the Product Owner who represents the stakeholders, and the Team which includes the developers.
>
> During each sprint, a 15-30 day period (length decided by the team), the team creates an increment of potential shippable (usable) software. The set of features that go into each sprint come from the product backlog, which is a prioritized set of high level requirements of work to be done. Which backlog items go into the sprint is determined during the sprint planning meeting. During this meeting the Product Owner informs the team of the items in the product backlog that he wants completed. The team then determines how much of this they can commit to complete during the next sprint. During the sprint, no one is able to change the sprint backlog, which means that the requirements are frozen for a sprint.
>
> There are several implementations of systems for managing the Scrum process which range from yellow stickers and white-boards to software packages. One of Scrum's biggest advantages is that it is very easy to learn and requires little effort to start using.
>
> `http://en.wikipedia.org/wiki/Scrum`Wikipedia 2/Sept/2008

Every day (or two, depending on preferences) the team meets for a very short round-up meeting called a *scrum*, where each team member states

- what they did since the last meeting
- what they plan to do next
- any blocking problems that they have

A key function of team leaders and managers who are present is to take responsibility for trying to remove blockers.

At sprint end a review and re-planning session adjusts the backlog and selects a new set of items for the next sprint. During sprints any new ideas or requirements for the product are added to the backlog, but **are not added** to the sprint – i.e. when a sprint has begun the task list is fixed for its duration. This means that developers can concentrate on one thing at a time (but because sprints are short it also means that managers can adjust the schedule quickly enough to cope with changing priorities).

Pictorially: see figure 2.1.

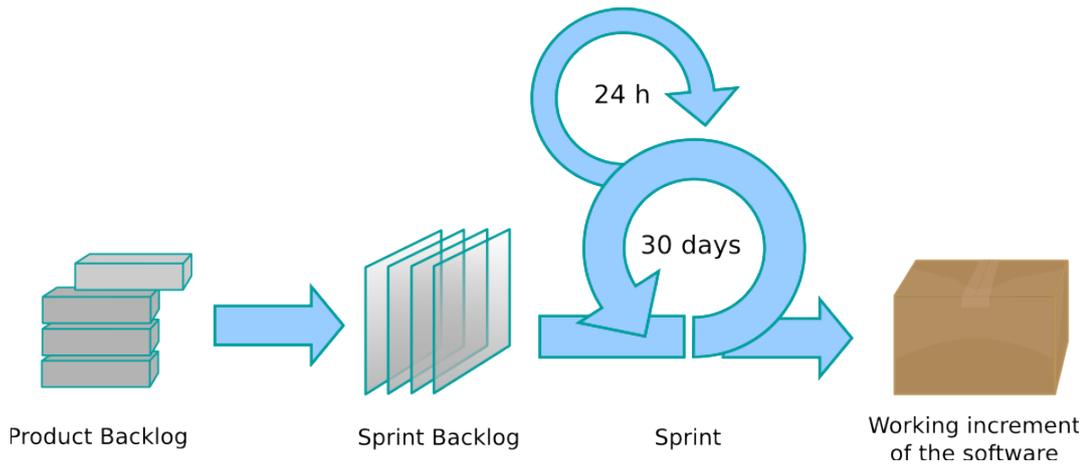

Figure 2.1: Scrum in pictures

## 2.2 Supporting the Programmer with XP

In 1999 Kent Beck wrote a book called *Extreme Programming Explained: Embrace Change* [Beck 00] that advocated dispensing with most design artefacts and implementing a programmer-centric division of labour. The philosophy was to accept the dynamism and complexity of the software context and to make a virtue out of a weakness by focussing on flexibility and the ability to change course at any point. Key techniques were to keep code quality high (to *refactor* at every opportunity, building libraries of reusable code instead of continually reinventing incrementally different solutions), maximise communication between programmers and between programmers and clients, and to test, test, and test some more as the main method of verifying functionality.

The original list of practises was:

**Small Releases:** Put a simple system into production quickly, then release new versions on a very short cycle.

**Metaphor:** Guide all development with a simple shared story of how the whole system works.

**Simple Design:** The system should be designed as simply as possible at any given moment. Extra complexity is removed as soon as it is discovered.

**Testing:** Programmers continually write unit tests, which must run flawlessly for development to continue. Customers write tests indicating that features are finished.

**Refactoring:** Programmers restructure the system without changing its behaviour to remove duplication, improve communication, simplify, or add flexibility.

- **Pair Programming:** All production code is written with two programmers at one workstation.
- **Collective Ownership:** Anyone can change code anywhere in the system at any time.
- **Continuous Integration:** Integrate and build the system many times a day, every time a task is completed.
- **40-hour Week:** Work no more than 40 hours a week as a rule. Never allow overtime for the second week in a row.
- **On-site Customer:** Include a real, live customer on the team, available full-time to answer questions.
- **Coding Standards:** Programmers write all code in accordance with rules emphasizing communication throughout the code.

[Beck 00, page 54]

Regarding not working too much overtime, we have refined the protocol as follows:

- on Monday we recover from the weekend
- on Tuesday we get ready to work
- on Wednesday we work
- on Thursday we recover from work
- on Friday we get ready for the weekend

This keeps most of the team happy, although when we first discussed the idea one member raised their hand and asked "Does this mean we're going to work *every* Wednesday?"

Next we look at the promise of the new methods in a research context.

# Chapter 3

# Agile Research

What of research? Here's how *not* to do it:

**Project Proposal:** We love each other. We can work so well together. We can hold workshops on Greek islands together. We will solve all the problems of AI that our predecessors were too stupid to manage.

**Analysis and Design:** Stop work entirely, for a period of reflection and recuperation following the stress of attending the kick-off meeting.

**Implementation:** Each developer partner tries to convince the others that program X that they just happen to have lying around on a dusty disk-drive meets the project objectives exactly and should form the centrepiece of the demonstrator.

**Integration and Testing:** The lead partner gets desperate and decides to hard-code the results for a small set of examples into the demonstrator, and have a fail-safe crash facility for unknown input ("well, you know, it's still a prototype...").

**Evaluation:** Everyone says how nice it is, how it solves all sorts of terribly hard problems, and how if we had another grant we could go on to transform information processing world-wide (or at least the European business travel industry).

[Cunningham & Bontcheva 05]

Seriously, there are significant risks hidden under the contractual carpet in modern ICT research, especially in shared-cost collaborative projects with many partners, many agendas, many pieces of background software that theoretically will be integrated and work smoothly together to demonstrate some radical new proposition regarding the possibility of the next high-tech revolution. For obvious reasons of financial probity the European Union variety of these projects begin with the authoring of a lengthy annex to the funding contract that specifies several acres of milestones, deliverables and other checkpoints. Conformance to the plans set out in this annex (the *description of work* in current parlance) is measured by periodic expert peer review in the normal (academic) manner.

This is all well and good, but from a software development perspective it does tend to encourage staying with the type of waterfall methods which have been shown to be less than perfect in dynamic environments. How to cope?

Our recommendations are to specify outputs as a series of iterations and adopt an agile process for the development of these prototype instances. The rest of the section summarises this process.

## 3.1 Process Summary

Reprising the above, XP is a way to keep focussed on user needs and to ensure stability through testing. Scrum is a way to organise regular deadlines and to encourage teamwork towards clearly defined objectives, while avoiding lock down to inflexible targets. Scrum is less about the code and more about how to split development time into manageable chunks, how to structure implementation iterations, and how to filter and prioritise ideas for new features and technical changes.

This section summarises an agile process for the GATE research team (`http://gate.ac.uk/`). The main elements of XP and Scrum that we have adopted are:

- Test-driven development.

- Deliver early, deliver often.

- Only document where necessary (minimise design artefacts because they go out of date or cost lots to maintain). Code comments plus a developer guide with some minimal notes on major design issues or central algorithms are a good baseline. (Note that user documentation is a different issue.)

- Maintain our library-based approach and refactor code into the frameworks in and around GATE. (Resist the temptation to reassure bureaucrats and marketing people of the novelty of the work by continually inventing new names.)

- Maintain our continuous integration suite with (minimally) nightly builds.

- Development done by self-organising teams.

- Development done on a bi-weekly or monthly cycle; no new work can be taken on by a team after a cycle starts.

- The list of features, functions, technologies, enhancements and bug fixes that would ideally be performed are prioritised and listed in a "backlog" which represents a snapshot of the ever-changing requirements and plan.

- For each development cycle pick the highest priority items to do next.

- There are no bad ideas for new features, only low priority features.

- Meet every 1 or 2 working days to briefly report: what's been done since last meeting; what is planned for next; what has been a barrier. These are called "scrums" and should last around 15 minutes.

The cycle:

- Develop a prioritised list of functionality, technology changes, bug fixes etc.

- Select a month's work (or a week or a fortnight, depending on external factors).

- Do a month's work (a "sprint").

- Review.

- Repeat.

During a sprint new work items cannot be assigned to the developers unless they are of an urgent, show-stopper nature. (New items that are identified are added to the backlog instead.) The team can reduce or increase the amount of work in the sprint if it needs to, but tries to avoid reductions if possible.

As new functionality becomes available it is rolled into the applications and user feedback sought.

Artefacts:

- Applications.

- Backlog. This lists all new and changed functionality that we've decided we need to develop to improve the applications. Items at the top of the list should be detailed enough and fine-grained enough to be implemented.

- Test suite. All the backlog items that the team produces should be associated with a set of tests.

- User guide. This has to contain enough information for users to understand what the team outputs and to exploit it effectively.

- Developer guide. This is written relatively informally as a set of notes that point into the code and test suite javadocs/java2html.

- Documentation process: the sequence is: vision; backlog to-do; backlog done, linked to developer or user documentation.

## 3.2 Programmer as Human

One of the nicest elements of the agile methods is the attention they pay to the social factors in software development. XP in particular has a humanist focus on developers that reflects the truism that people work well when they're happy and feel that they are doing useful work. There's a down side, which is that a significant effort of will is required to establish a process, and the overhead involved will always feel unwelcome to some (us computer types are indeed human, but are not always the most naturally communicative members of the species). This means that agile processes need active champions within the organisations that use them, especially in the early stages, and that judicious deployment of foot massage, alcohol, chocolate, or other health foods may also be considered necessary, depending on context.

# References

[Beck 00]
> K. Beck. *eXtreme Programming eXplained*. Addison-Wesley, Upper Saddle River, NJ, USA, 2000.

[Booch 94]
> G. Booch. *Object-Oriented Analysis and Design 2nd Edn.* Benjamin/Cummings, 1994.

[Brand 94]
> S. Brand. *How Buildings Learn*. Penguin, London, 1994.

[Cunningham & Bontcheva 05]
> H. Cunningham and K. Bontcheva. Computational Language Systems, Architectures. *Encyclopedia of Language and Linguistics, 2nd Edition*, pages 733–752, 2005.

[Frederick P. Brooks 78]
> J. Frederick P. Brooks. *The Mythical Man-Month: Essays on Software Engineering*. Addison-Wesley Longman Publishing Co., Inc., Boston, MA, USA, 1978.

[Schwaber & Beedle 01]
> K. Schwaber and M. Beedle. *Agile Software Development with Scrum*. Prentice Hall PTR, Upper Saddle River, NJ, USA, 2001.

[Yourdon 89]
> E. Yourdon. *Modern Structured Analysis*. Prentice Hall, New York, 1989.